\documentclass[lettersize,journal]{IEEEtran}
\usepackage{amsmath,amssymb,amsfonts}
\usepackage{algorithmic}
\usepackage{algorithm}
\usepackage{array}
\usepackage[font=scriptsize]{subfig}
\usepackage{textcomp}
\usepackage{stfloats}
\usepackage{url}
\usepackage{verbatim}
\usepackage{graphicx}
\usepackage{cite}
\usepackage{color}
\usepackage{makecell}
\begin{document}

\title{Federated Neural Radiance Field for Distributed Intelligence}

\author{Yintian Zhang, Ziyu Shao \\
School of Information Science and Technology, ShanghaiTech University, Shanghai, China\\
Email: \{zhangyt, shaozy\}@shanghaitech.edu.cn
\thanks{Placeholder for Thanks}
\thanks{Manuscript received Month XX, 20XX; revised Month XX, 20XX.}}

\markboth{Journal of \LaTeX\ Class Files,~Vol.~XX, No.~XX, Month~2024}%
{Shell \MakeLowercase{\textit{et al.}}: A Sample Article Using IEEEtran.cls for IEEE Journals}


\maketitle

\begin{abstract}
    Novel view synthesis (NVS) is an important technology for many AR and VR applications. The recently proposed Neural Radiance Field (NeRF) approach has demonstrated superior performance on NVS tasks, and has been applied to other related fields. However, certain application scenarios with distributed data storage may pose challenges on acquiring training images for the NeRF approach, due to strict regulations and privacy concerns. In order to overcome this challenge, we focus on FedNeRF, a federated learning (FL) based NeRF approach that utilizes images available at different data owners while preserving data privacy. 

    In this paper, we first construct a resource-rich and functionally diverse federated learning testbed. Then, we deploy FedNeRF algorithm in such a practical FL system, and conduct FedNeRF experiments with partial client selection. It is expected that the studies of the FedNeRF approach presented in this paper will be helpful to facilitate future applications of NeRF approach in distributed data storage scenarios.
\end{abstract}


\section{Introduction}
\IEEEPARstart{S}{ynthesizing} novel views of a scene from a set of captured images in discrete views is a long-standing problem in the computer vision field, and a prerequisite to many AR and VR applications. Recently proposed new techniques, as known as \textit{neural rendering}, have made significant advancements in this field. These techniques use learning-based modules to represent the scene, train the modules to reconstruct observed images, and then synthesize novel views of the scene with well-trained modules. The Neural Radiance Fields (NeRF) approach \cite{mildenhall2020nerf} is a representative work among these neural rendering techniques, and it demonstrates a heretofore unprecedented level of fidelity on a range of challenging scenes. Consequently, a multitude of researchers have pursued NeRF research and proposed numerous improved algorithms and extended applications in diverse fields and scenarios.

However, NeRF and its subsequent approaches encounter challenges in certain application scenarios where data storage is distributed. For example, the \textit{NeRF in the Wild} approach \cite{martin2021nerf} proposes to gather images of the same scene captured by different user devices in different conditions together, to train the neural network representing the scene. But the users may not be willing to upload the images because the images may contain users' privacy. Another example is the medical image processing scenarios. The extended applications of the NeRF approach in the medical image processing field \cite{wysocki2023ultra} \cite{corona2022mednerf} may be troubled by data silos problems. Because it is usually complex to transfer and share medical data between different medical institutions, due to strict regulations and sensitivity of medical data \cite{liu2018fadl}. In this situation, the images data stored in different medical institutions cannot be maximally utilized for training a high-quality model.
\begin{figure}[t]
    \includegraphics[width=1.0\linewidth]{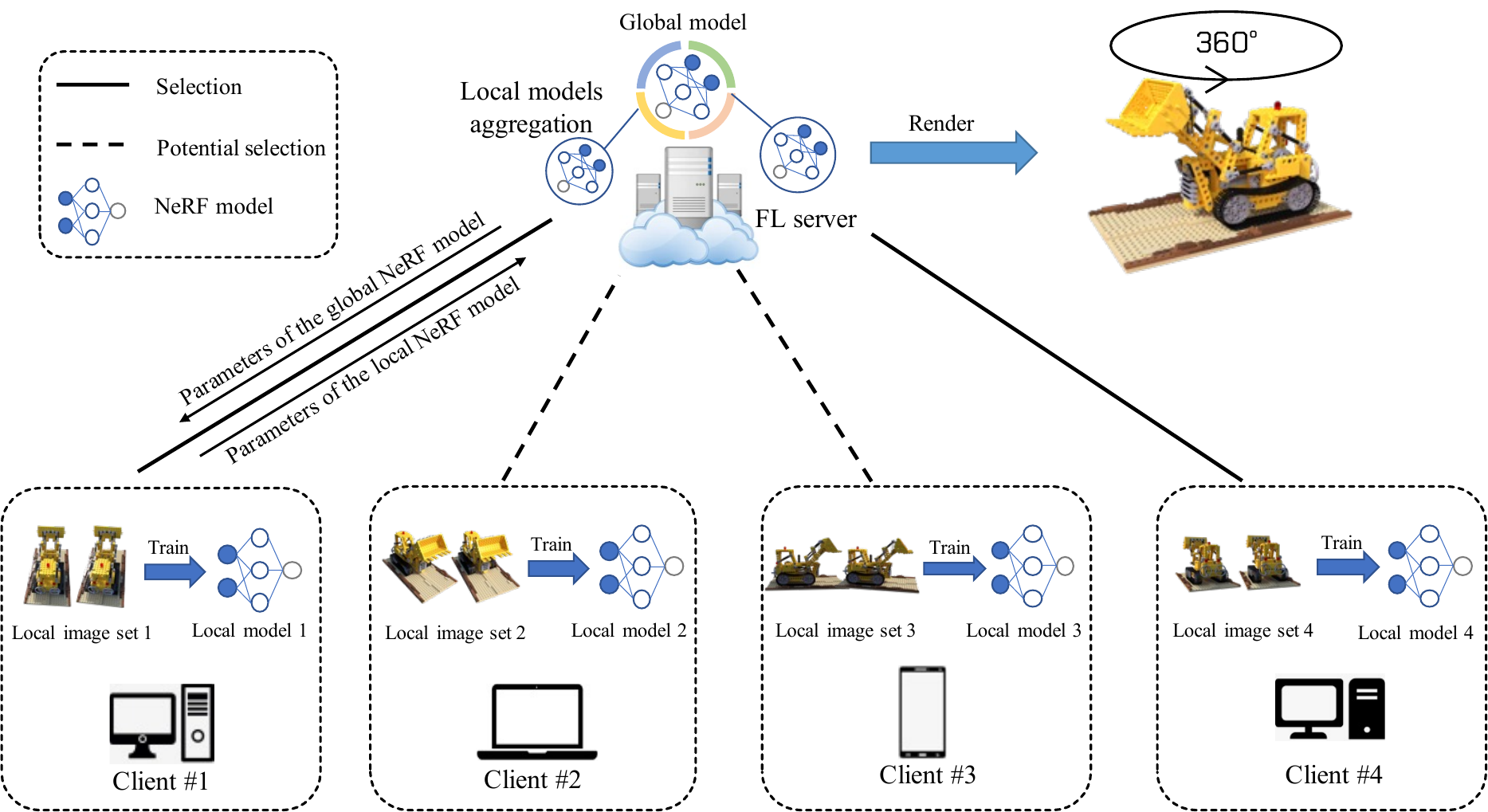}
    \captionsetup{font={scriptsize}}
    \caption{\textbf{An overview of the proposed FedNeRF approach}. Through several rounds of communication between the clients and the FL server, the collaboratively trained global NeRF model can learn omnidirectional scene information in a data privacy-preserving manner.}
    \label{fig:system}
\end{figure} 
The federated learning (FL) framework \cite{lim2020federated} is able to address such challenges by allowing collaborative and distributed training of deep learning-based methods, without raw training data transferring between the participants. Therefore, we focus on incorporating the NeRF approach into the FL framework, aiming at facilitating the applications of the NeRF approach in distributed data storage scenarios. This incorporating is referred to as the FedNeRF approach, illustrated as Figure \ref{fig:system}. 

In this paper, we first construct a federated learning testbed. Our testbed is equipped with a diverse set of wired and wireless computation nodes, and it boasts rich computational resources at multiple performance levels. Our testbed also has many additional functionalities for simulations beyond its core usage of facilitating FL tasks. With this FL testbed, we further conduct experiments the FedNeRF approach with partial client selection. The dynamic client selection algorithm we design achieves a trade-off effect between reducing transmission time overhead and improving model performance level.

\begin{figure*}[ht]
    \centering
    \captionsetup{font={scriptsize}} 
    \includegraphics[width=0.8\linewidth]{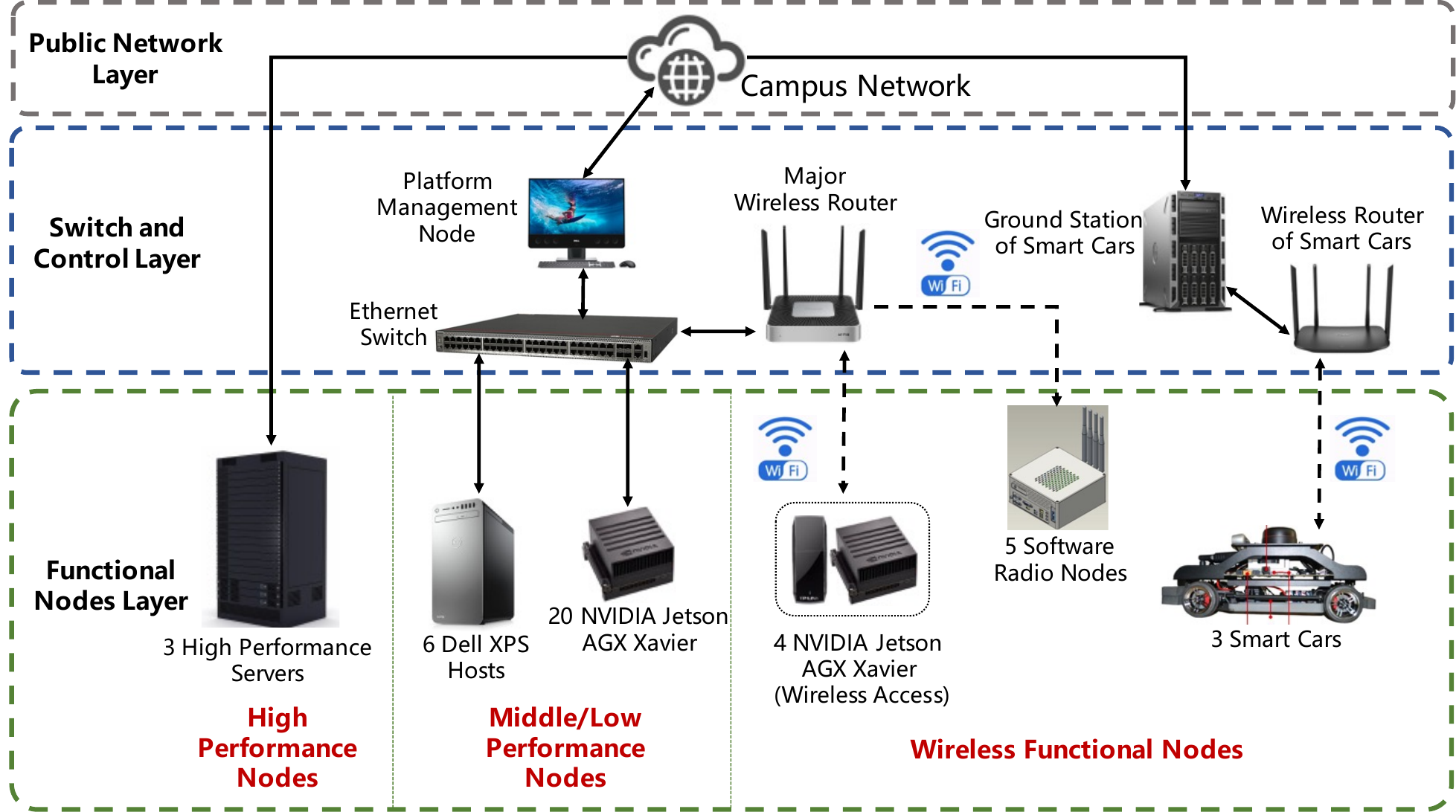}
    \caption{The structure of the federated learning testbed}
    \label{fig:testbed_struct}
\end{figure*}

\section{The Federated Learning Testbed}

The structure of the federated learning testbed is shown as Figure.\ref{fig:testbed_struct}. The structure of this FL testbed contains the following three layers. 

\begin{itemize}
    \item \textbf{Public Network Layer}. All nodes within this testbed ultimately achieve inter-connectivity and access to public internet resources through our campus network. All high-performance servers in the testbed are equipped with static IP addresses of the campus network, which directly meet the testbed's requirement for stable device addressing upon connection to the network. However, other devices require access to the campus network through the switch and control layer.
    \item \textbf{Switch and Control Layer}.The purpose of the equipments within the switch and control layer is to facilitate the connection of nodes within this testbed that do not possess static IPs to the public network. It also enables port forwarding to each device while providing unified management and control. The switch and control layer offers two ways of access for each node: wired Ethernet and wireless Wi-Fi.
    \item \textbf{Functional Nodes Layer}. The nodes in the functional nodes layer serve as the hardwares that embody various practical functionalities of the FL testbed. These functionalities will be described in the next paragraph. 
\end{itemize}

The FL testbed contains the following subsystems providing practical functionalities for experiments related with federated learning. 
\begin{itemize}
    \item \textbf{High Performance Nodes Cluster}. This cluster comprises 3 high-performance servers, equipped with 5 high-capacity independent GPUs totally. The cluster is primarily utilized for deploying simulation experiments that require substantial computational resources. It can be employed for simulations in the research of deep learning algorithms with high computational demands, as well as for single-node pseudo-distributed simulations of federated learning algorithms.
    \item \textbf{Middle/Low Performance Nodes Cluster}. This cluster consists of 8 hosts and 20 AI development boards. Each host is equipped with an independent GPU, while each development board is equipped with a GPU specifically designed for embedded AI devices. This cluster is primarily used to provide a sufficient number of deployment nodes for FL training tasks that have moderate computational requirements.
    \item \textbf{Wireless Nodes Cluster}. The cluster is composed of 4 AI development board terminals with wireless access, and 3 smart car terminals. Each node connects to its corresponding wireless router via Wi-Fi, thereby integrating into the testbed. This cluster is primarily designed to provide a fundamental simulation environment for the deployment of FL algorithms and applications in wireless network settings.
    \item \textbf{Traffic Scenarios Simulation Subsystem}. In this testbed, the 3 smart cars primarily play two roles. First, they act as low-performance wireless terminals to conduct federated learning experimental tasks. Second, they leverage their own computational capabilities and various peripherals to implement autonomous line-following driving functions, thereby simulating traffic scenarios. The ground station of the smart cars controls the autonomous driving functions of the cars. These two functionalities collectively ensure that the smart cars can provide a simulation environment for the deployment of federated learning algorithms and applications in scenarios where computational nodes are mobile.
    \item \textbf{Software Radio Nodes Cluster}. This cluster consists of 5 software-defined radio (SDR) nodes, which are utilized for conducting more in-depth simulations of the wireless physical layer environment that may be involved in federated learning experiments. Inside the node chassis, the computer motherboard is connected to a universal software radio peripheral (USRP) device motherboard via a USB cable. Researchers can develop and deploy software-defined radio applications in a one-stop manner on these SDR nodes, and they can conveniently transmit the execution results of the program in real-time to other nodes within the testbed via the wireless network cards equipped on these nodes.
\end{itemize}

\section{Conducting FedNeRF Tasks in the FL Testbed}\label{sec:fednerf_real_impl}
\begin{figure}[ht]
    \centering
    \captionsetup{font={scriptsize}} 
    \includegraphics[width=0.9\linewidth]{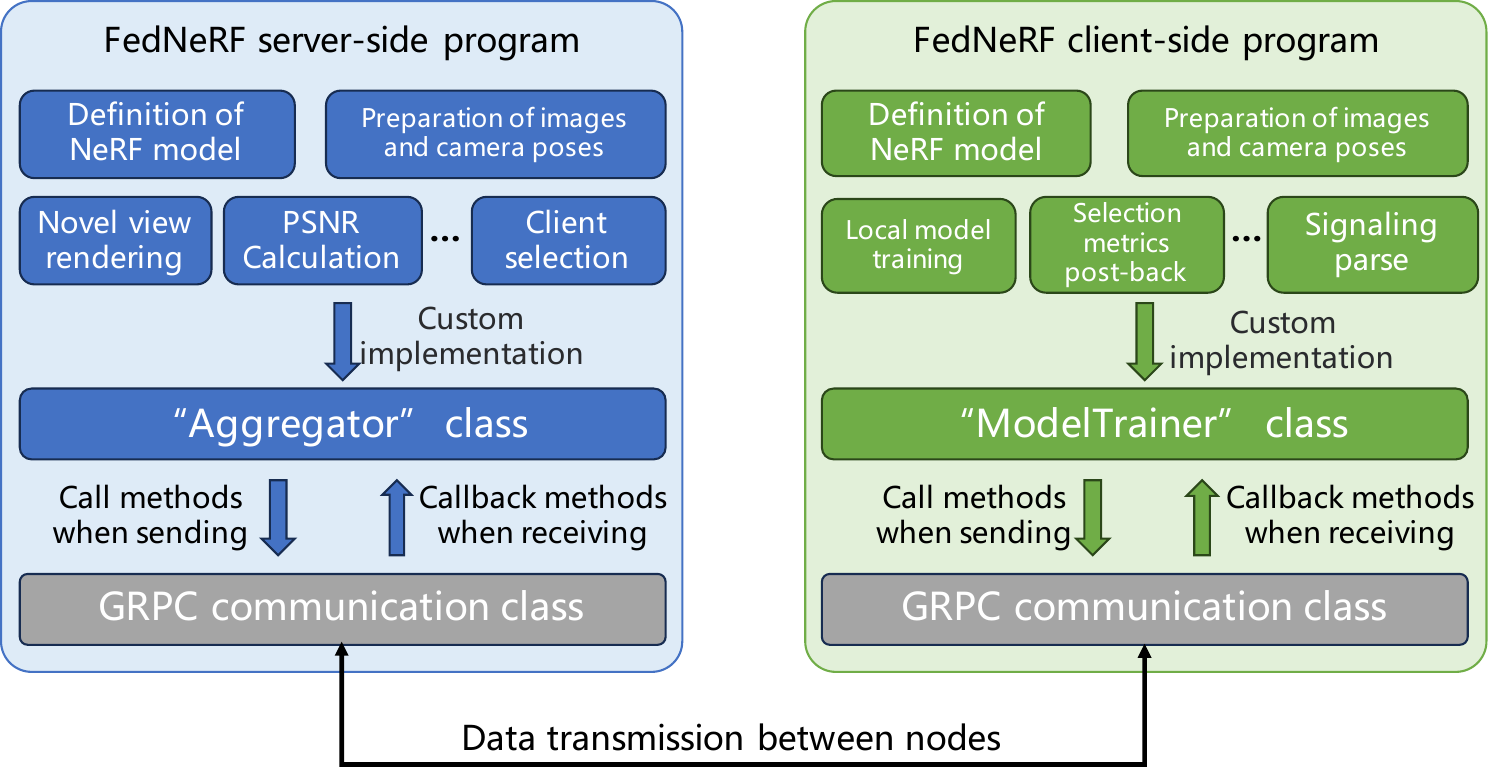}
    \caption{The implementation of FedNeRF in the FL testbed}
    \label{fig:testbed_coding}
\end{figure}

In this section, we will describe how the FedNeRF task is implemented and deployed in this testbed. Indeed, the process for deploying other types of FL tasks on this universal testbed is essentially consistent. 

Initially, all the computational nodes within this FL testbed are capable of hosting and executing the experimental programs related to FedNeRF research. Before deploying the experimental programs, researchers should delineate the scope of computational node resources required for the experiment, which involves determining which node in the testbed will act as the FL server and which nodes will act as client devices.

After defining the scope of computational node resources, researchers can proceed to implement the server-side and client-side programs corresponding to the FedNeRF method. In this testbed, the experimental programs of the FedNeRF method are implemented based on the FedML federated learning development framework \cite{he2020fedml}, as depicted in Figure \ref{fig:testbed_coding}. When implementing the FedNeRF method, researchers should first implement the relevant code for deep learning model definition, training data loading, and model training algorithms, following the prescribed methods within the FedML framework. 

After completing the coding of the custom parts, researchers need to inherit the ``Aggregator'' and ``Model Trainer'', two core functional superclasses provided by the FedML framework, in their experimental program code. By overriding the relevant methods in these classes with the aforementioned custom methods, researchers can implement both the server-side and client-side programs. As for the communication between nodes and the control for the FL process, these functions have already been implemented in the underlying code of the FedML framework.

Once the experimental programs for the server and clients have been implemented, researchers can deploy them to the designated computational nodes. They should also specify the device ID for each client node within the testbed in the server-side program and the server node's device ID in the client-side program. After launching the experimental programs, the client-side program will locate the server node's IP address in the testbed's configuration file based on the device ID, and proactively initiate a handshake with the server to establish a connection.

Finally, researchers can employ automated scripts and SSH remote connections to start the FedNeRF experimental programs on each node. After the experiments are completed, they can log in to each node to check the experiment results.

\section{Performance of FedNeRF with Client Selection in FL Testbed}\label{sec:fednerf_select_channel_real}
In this section, we will deploy the FedNeRF training with dynamic client selection algorithm on actual wireless devices, and conduct an effectiveness verification experiment in the FL testbed. 
We use the ground station server of the smart cars in the FL testbed to perform the function of the FL server, and also deploy four wireless AI development board terminals in the testbed to fulfill the role of clients. The server node is equipped with an NVIDIA GeForce RTX 3070Ti GPU, and the wireless terminals are NVIDIA Jetson AGX Xavier development boards equipped with USB wireless network cards. The server connects to each client via a wireless router using Wi-Fi5 protocol, to facilitate the transmission of model parameters. The model of the wireless router equipped by the server is TL-WDR5620. The wireless network card on each client can measure the current Received Signal Strength Indicator (RSSI) in real time, with the value in the range of 0-100. The larger this indicator, the stronger the signal at the receiver, and the better the channel quality. The implementation of the experimental procedure is based on the FedML framework, as it has been introduced in Section \ref{sec:fednerf_real_impl}. 

In the experiments of this section, the four wireless clients are grouped in pairs, with one group placed closer to the wireless router and the other group placed further away from the wireless router. The differences in the distances from each client to the wireless router leads to variations in the measured RSSI, and consequently the differences in wireless channel quality. The measured RSSI of the four clients are shown in Table \ref{tab:device_rssi}. Based on the RSSI measurements of each client, we map the RSSI within different value ranges to channel quality levels according to Table \ref{tab:signal_level_to_quality_level}.

\begin{table}[htb]
    \centering
    \begin{tabular}{|c|c|c|c|c|}
      \hline
      \textbf{Client ID}         & 1  & 2   & 3  & 4  \\ \hline
      \textbf{RSSI Measurement} & 50 & 41  & 66  & 73 \\ \hline
    \end{tabular}
    \caption{The table of measured RSSI of each client}
    \label{tab:device_rssi}
\end{table}

\begin{table}[htb]
    \centering
    \begin{tabular}{cc}
      \hline
      \textbf{RSSI Range}        & \textbf{Channel Quality Level($z$)} \\ \hline
      RSSI $\leq$ 50          & 1                    \\ 
      50 $<$ RSSI $\leq$ 60     & 2                    \\ 
      60 $<$ RSSI $\leq$ 70     & 3                    \\ 
      RSSI $>$ 70               & 4                    \\ \hline       
    \end{tabular}
    \caption{The table of RSSI range mapping to channel quality level}
    \label{tab:signal_level_to_quality_level}
\end{table}

In the experiments of this section, the federated learning server selects two out of four client devices in each round. The the local training datasets on each client contains 4 scene images, with 1 view as the novel view for testing. In the experiments of this section, the number of training rounds is 400, with 100 iterations of local training each round.

\begin{figure*}[htb]
    \centering
    \subfloat[\centering{Curves of global NeRF model performance level \newline with different $Q$ values}]{
      \includegraphics[width=0.4\linewidth]{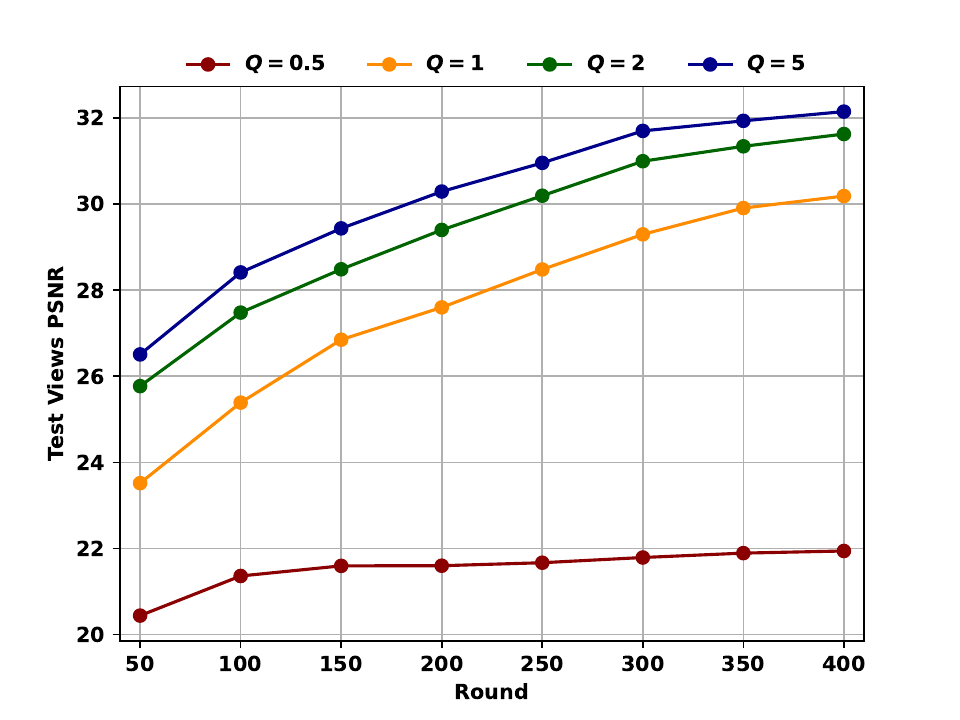}
      \label{fig:real_wireless_q_psnr}
    }
    \subfloat[\centering{Curves of model transmission rate level \newline with different $Q$ values}]{
      \includegraphics[width=0.4\linewidth]{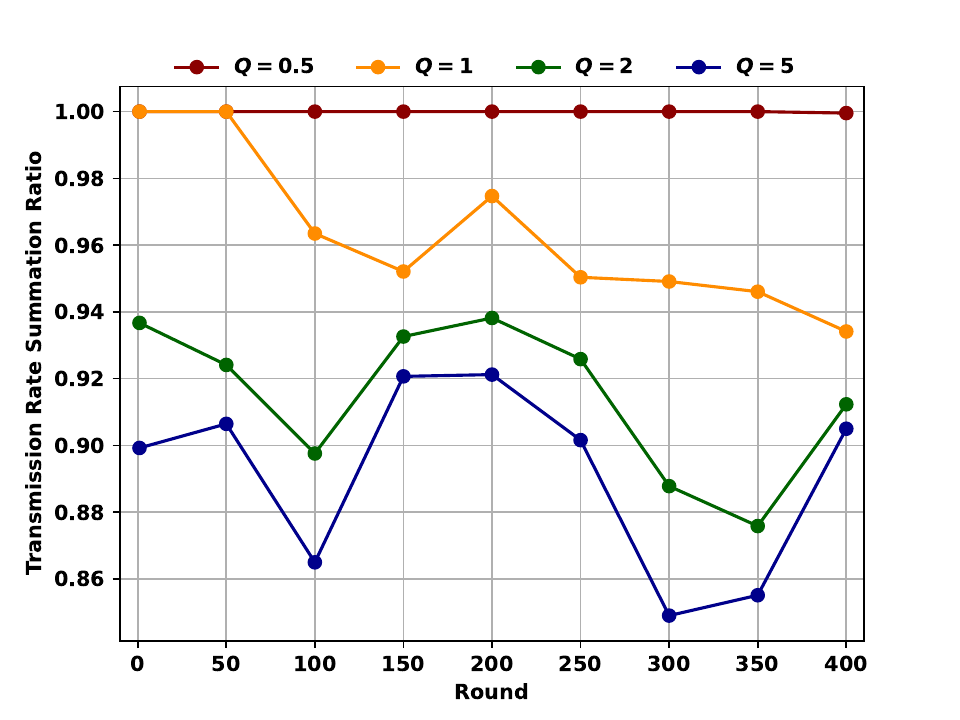}
      \label{fig:real_wireless_q_capacity}
    }
    \caption{Experiment results of the FedNeRF-CS on real wireless devices}
    \label{fig:real_wireless}
\end{figure*}

\begin{table}[htb]
    \centering
    \begin{tabular}{|c|c|c|c|c|}
      \hline
      \textbf{Client ID}         & 1  & 2   & 3  & 4  \\ \hline
      \textbf{Transmission Rate (Mbit/s)} & 217.48 & 197.18  & 270.43  & 305.81 \\ \hline
    \end{tabular}
    \caption{The measured transmission rates from the server to each client}
    \label{tab:device_trans_rate}
\end{table}

We conduct experiments with different values of hyper-parameter $Q$, in order to demonstrate its trade-off effects. We will present variation in model performance levels and model transmission rate levels during the training process. The model performance level is still characterized by the mean PSNR of the NVS results. As for the model transmission rate level, we directly measure the downlink data transmission rate from the server to each client. The transmission rate is measured through conducting ``iperf'' command on the FL server, in the units of Mbit/s. We conduct 80 times of  measurements of the downlink transmission rate from the FL server to each client, and the average of the speed measurement results for each client are shown in Table \ref{tab:device_trans_rate}. Then, we replace the channel capacity metric with the transmission rate metric measured in each round, and derive the metric of the ``ratio of selected client transmission rate summation''. This ratio metric is adopted to reflect the overall level of model transmission rates in the current round. 

The experiments results with different values of the hyper-parameter $Q$ are shown in Figure \ref{fig:real_wireless}. Similar to the single-node simulation, in Figure \ref{fig:real_wireless_q_capacity}, the the metric of the ratio of selected client transmission rate summation corresponding to each round is the average of the ratio metric for a total of 10 rounds before and after that round. The conclusions revealed by the results in Figure \ref{fig:real_wireless} are quite obvious. Adjusting the hyper-parameter $Q$ indeed achieves a trade-off effect between improving model transmission rate (i.e., reducing transmission time overhead) and improving model performance level. Therefore, the experimental results in this section further prove the effectiveness of the FedNeRF-CS algorithm considering channel qualities in practical systems.

\section{Conclusion}
Our study focuses on FedNeRF, an FL-based approach to leverage distributed stored image data for the neural radiance field training task in a privacy-preserving manner. We have implemented a federated learning testbed, and have deployed FedNeRF experiments with partial client selection in such a practical system. The studies of the FedNeRF approach will be helpful to facilitate future applications of NeRF approach in distributed data storage scenarios.


\bibliographystyle{./IEEEtran}
\bibliography{./FEDNERF_DRAFT_ARXIV}
 


\end{document}